\documentclass[twoside]{dis04}
\def\runauthor{S.~Sch\"atzel}
\def\shorttitle{Diffractive final states and QCD predictions}

\newcommand{\xgamjets}{%
\ensuremath{%
{x_\gamma^{\rm jets}}}}
\newcommand{\zpomjets}{%
\ensuremath{%
{z_\pom^{\rm jets}}}}
\newcommand{\xgam}{%
\ensuremath{%
{x_\gamma}}}
\newcommand{\figref}[1]{Fig.~\ref{#1}}
\newcommand{\pom}{{I\hspace{-0.5ex} P}}
\newcommand{\zpom}{%
\ensuremath{%
{z_\pom}}}
\def\publ{}
\newcommand{\mjj}{%
\ensuremath{%
{M_{12}}}}
\newcommand{\xpom}{%
\ensuremath{%
{x_\pom}}}
\newcommand{\etjetstar}{%
\ensuremath{%
{E_T^{\rm *,jet}}}}
\newcommand{\etjet}{%
\ensuremath{%
{E_T^{\rm jet}}}}
\newcommand{\ptjetone}{%
\ensuremath{%
{p_T^{\rm jet1}}}}
\begin{document}
\title{Comparison of diffractive final states \\
       with LO and NLO QCD predictions}

\author{S.~Sch\"atzel\footnote{On behalf of the H1 Collaboration}}
\address{Physikalisches Institut der Universit\"at Heidelberg \\
69120 Heidelberg, Germany \\
E-mail: schaetzel@physi.uni-heidelberg.de}

\maketitle

\abstracts{%
\noindent Measurements of hard diffractive final states performed with the H1
experiment at HERA are presented
and confronted with LO and NLO QCD predictions based on 
diffractive parton densities to test QCD factorisation in diffraction.
}

\section{Introduction}
\begin{figure}[!thb]
\vspace*{6.0cm}
\begin{picture}(20,10)
\includegraphics{qqbarg.eps}
\includegraphics{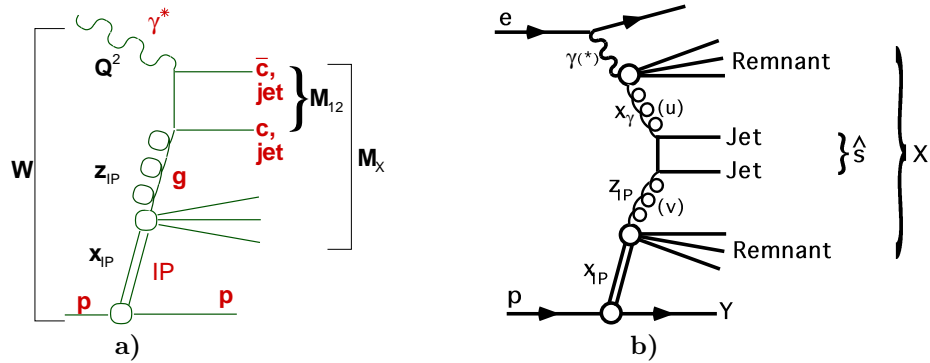}
\put(55,55){\textbf{a)}}
\put(250,55){\textbf{b)}}
\end{picture}
\vspace*{-2.2cm}
\caption[]{a) Direct photon process (photon-gluon fusion) and b) resolved photon process.}
\label{fig:bgf}
\vspace*{-0.5cm}
\end{figure}

The understanding of the production mechanism of diffractive processes
in high energy particle physics has remained
one of the most serious challenges in Quantum Chromodynamics (QCD),
the theory of strong interactions.
QCD predicts that the cross section
for diffractive deep-inelastic
electron-proton scattering (DDIS) factorises into universal diffractive parton
densities (DPDFs) of the proton and process-dependent hard scattering
coefficients~\cite{collins} (QCD factorisation).
Leading order
(LO) and next-to-leading order (NLO) DPDFs have been determined 
from DGLAP QCD fits to inclusive DDIS measurements by the H1
collaboration~\cite{h1f2d94,h1f2d97} and have been found to be
dominated by the gluon which carries $\approx 75\%$ of the momentum.
Diffractive dijet and D$^*$ meson (heavy quark) production are
directly sensitive to
the diffractive gluon through the photon-gluon fusion production mechanism
(\figref{fig:bgf}a) and are used to test factorisation.
A photon of virtuality $Q^2$ undergoes a hard scatter with
a diffractive gluon forming a $q\bar{q}$ pair. The centre-of-mass
energy of the hard scattering process is labelled \mjj.
The gluon carries a fraction \zpom{} of the momentum of the
diffractive exchange which itself carries a fraction \xpom{} 
of the proton momentum.
The $\gamma p$ centre-of-mass energy is denoted by $W$, and $M_X$
labels the mass of the diffractive system.
The inelasticity variable $y$ is given by $y\,s = W^2 + Q^2$, in which
$s$ is the squared $ep$ centre-of-mass energy.
The cross sections are also shown as a function of the pseudorapidity
$\eta$ of the jets and D$^*$ mesons.
Contributions to the cross section also occur through the process
depicted in \figref{fig:bgf}b
where the photon develops hadronic structure of which a single
parton with photon momentum fraction $\xgam$ undergoes 
the hard scatter (``resolved'' photon process).
In photoproduction ($Q^2 \approx 0$), this process contributes significantly
whereas in DIS it is suppressed due to the large photon virtuality.

\vspace*{-0.2cm}
\section{H1 diffractive parton distributions}
The H1 collaboration has determined diffractive parton densities
from NLO and LO DGLAP QCD fits to inclusive DDIS measurements. 
The latest fit to the most recent available data has been presented as
a preliminary result in~\cite{h1f2d97}. This fit is
referred to as `H1 2002 fit.' Earlier fits have been presented 
in~\cite{h1f2d94}, where the fit which gave the best description of
the inclusive process is referred to as `H1 fit 2.'

\vspace*{-0.2cm}
\section{Diffractive dijet production in DIS}
\begin{figure}[!thb]
\vspace*{6.7cm}
\begin{picture}(20,10)
\includegraphics{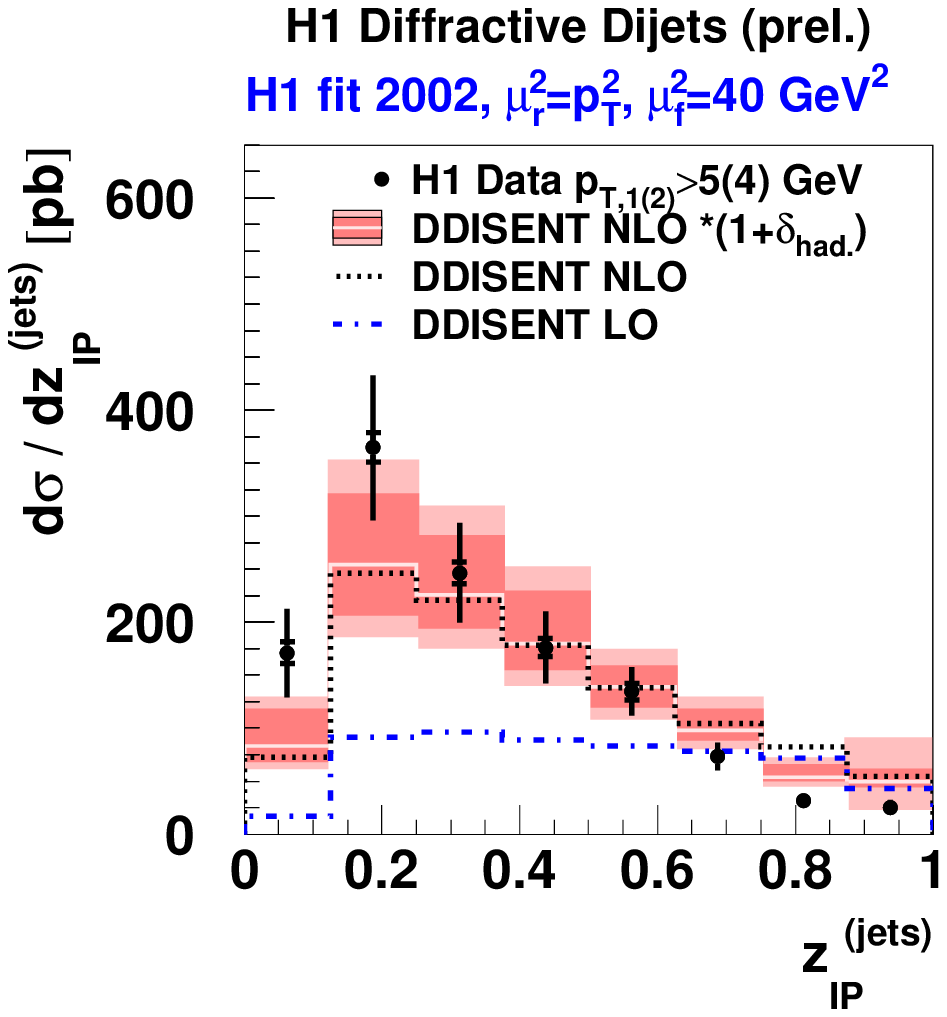}
\includegraphics{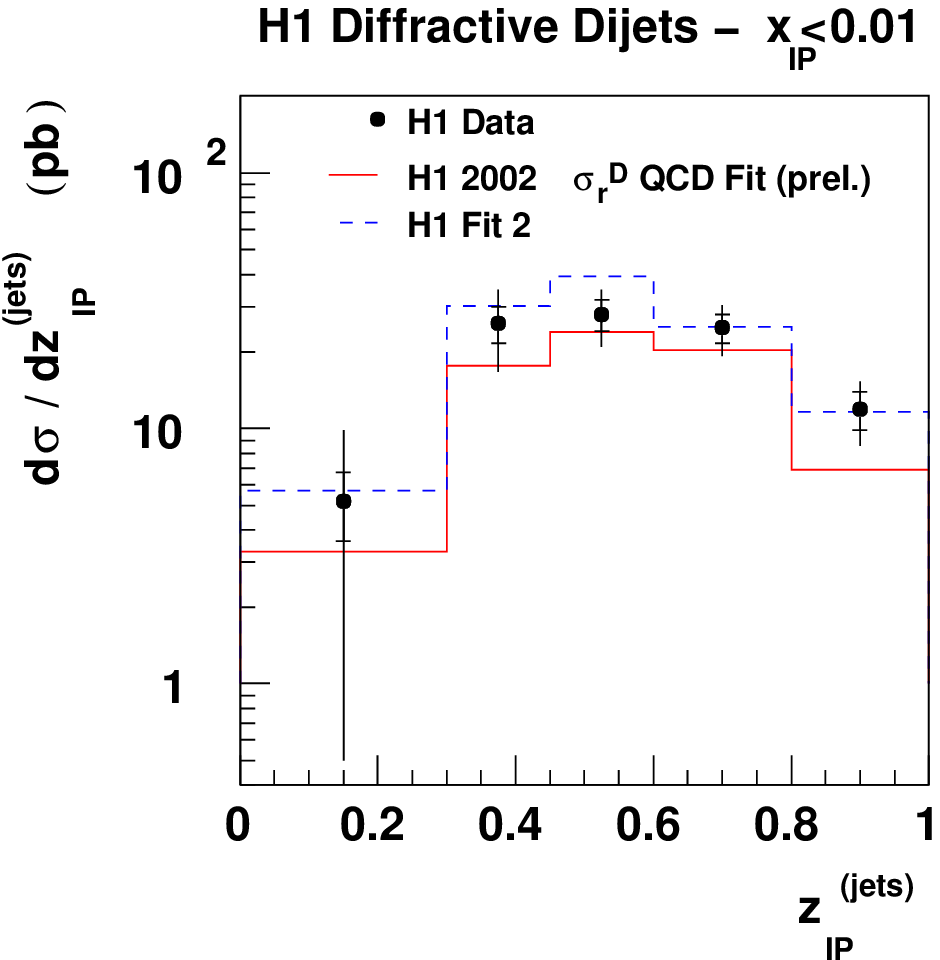}
\put(85,64){\textbf{a)}}
\put(260,64){\textbf{b)}}
\end{picture}
\vspace*{-2.5cm}
\caption[]{Diffractive DIS dijet cross section as a function of
the estimator $\zpomjets$ of the parton momentum fraction of the diffractive
exchange entering the hard scatter. a) For 
$E_T^{\rm *,jet1(2)}>5(4)$~GeV and $\xpom<0.05$ 
compared with NLO and LO DISENT based on the `H1 2002 fit' DPDFs.
b) For $E_T^{\rm *,jet1,2}>4$~GeV, $\xpom<0.01$
compared with the LO Monte Carlo RAPGAP with parton showers based on LO DPDFs.}
\label{fig:diszpom}
\vspace*{-0.7cm}
\end{figure}

\begin{figure}[!thb]
\vspace*{10.7cm}
\begin{center}
\includegraphics{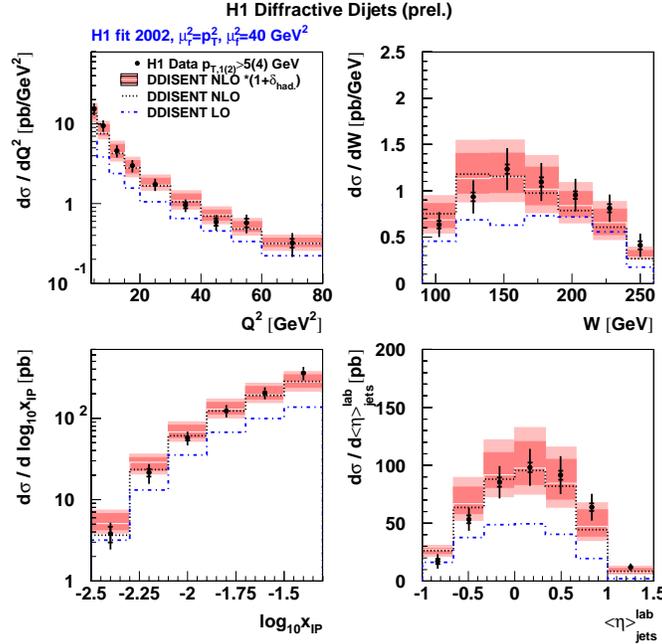}
\vspace*{-2.5cm}
\caption[]{Diffractive DIS dijet cross section as a function of
various variables compared with NLO and LO predictions of the DISENT
program based on the `H1 2002 fit' DPDFs.}
\label{fig:disother}
\end{center}
\vspace*{-0.8cm}
\end{figure}

Cross sections for diffractive dijets production in the kinematic
range $Q^2>4$~GeV$^2$, $\etjetstar(1,2)>4$~GeV, $\xpom<0.05$ and
$\xpom<0.01$ have been
measured by H1 in~\cite{disjets} using a cone jet algorithm. 
To facilitate comparisons with NLO calculations
the measured distributions have been
corrected to asymmetric cuts $\etjetstar(1)>5$~GeV and
$\etjetstar(2)>4$~GeV using Monte Carlo generated events~\cite{nloeps}.
To obtain NLO predictions, the NLO versions of the `H1 2002 fit' DPDFs 
were interfaced to 
the \mbox{DISENT} program~\cite{disent} as suggested in~\cite{hautmann}.
The renormalisation and factorisation scales were set to the average
$p_T$ of the two highest $p_T$ partons. The strong coupling $\alpha_s$
is calculated from $\Lambda^{\overline{\rm MS}}_{n=4}=200$~MeV. 
The same $\alpha_s$ is used in the DGLAP evolution of the DPDFs.
The NLO parton jet cross sections have been corrected for
hadronisation effects using the Monte Carlo program
RAPGAP~\cite{rapgap} with parton showers and Lund string
fragmentation.
Comparisons of the NLO and LO prediction with the measurement are 
shown in \figref{fig:diszpom}a and \ref{fig:disother}.
The NLO correction amounts to more than a factor 2 on average
and is decreasing with $\etjetstar$ (not shown) and $Q^2$. 
The inner band around the
NLO results indicate the $\approx 20\%$ uncertainty resulting
from a variation of
the renormalisation scale by factors 0.5 and 2. The outer band
includes a $\approx 10\%$ hadronisation uncertainty added linearly.
The uncertainty in the diffractive gluon distribution is not shown.
For $\zpom>0.7$ it is larger than 50\%.
Within the experimental and theoretical uncertainties, 
the cross section is well
described by the NLO calculation assuming QCD factorisation.

In \figref{fig:diszpom}b, 
a comparison is shown of the cross section in the range 
$E_T^{\rm *,jet1,2}>4$~GeV and $\xpom<0.01$ with
LO predictions of the RAPGAP Monte Carlo
program based on LO DPDFs.
Higher-order effects are modelled by using parton showers.
The `H1 fit 2' prediction describes 
the measurement well.
The new `H1 2002 fit' DPDFs lead to a $\approx
25\%$ lower cross section. The difference is of the order of the
uncertainty arising from that of the gluon distribution.

\section{Diffractive production of D$^*$ mesons in DIS}
Cross sections for diffractive D$^*$ meson production in DIS have been measured
by H1 in~\cite{dstar} in the kinematic range
$2<Q^2<80$~GeV$^2$, $\xpom<0.04$, and $p_T^{D^*} > 2$~GeV.
QCD calculations based on the H1 DPDFs have been performed at LO and
NLO using the diffractive extension~\cite{dhvqdis} of the 
HVQDIS program~\cite{hvqdis}. The
renormalisation and factorisation scales were set to
$\mu^2=Q^2+4m_c^2$. Other parameter values used include the charm
quark mass $m_c=1.5$~GeV, the hadronisation fraction \mbox{$f(c\rightarrow
D^*)=0.233$}, and $\varepsilon=0.078$ for the Peterson fragmentation
function~\cite{peterson}.

\begin{figure}[!thb]
\vspace*{10.8cm}
\begin{center}
\includegraphics{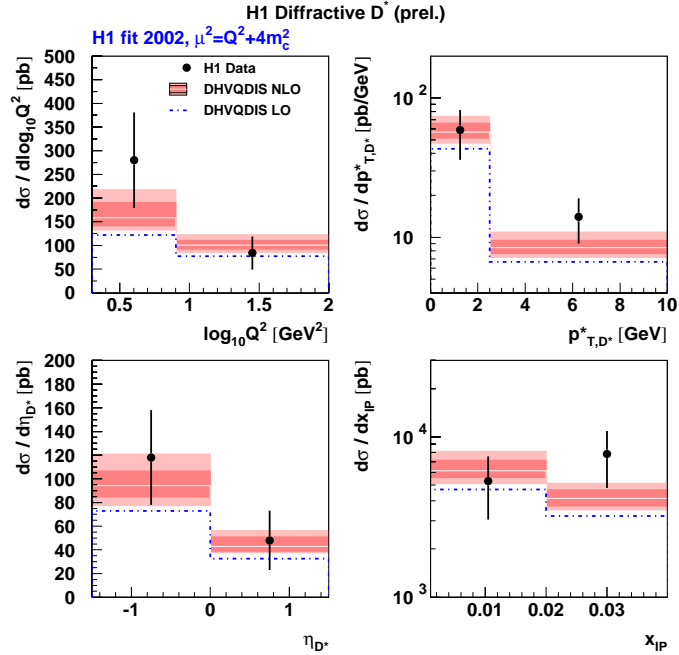}
\vspace*{-2.5cm}
\caption[]{Diffractive DIS D$^*$ cross section as a function of
various variables compared with NLO and LO predictions of the HVQDIS
program extended for diffraction and using the `H1 2002 fit' DPDFs.}
\label{fig:dstar}
\end{center}
\vspace*{-1.5cm}
\end{figure}

Comparisons of LO and NLO predictions with measured cross sections are shown in
\figref{fig:dstar}. The NLO correction amounts to a factor $\approx
1.3$ on average.
The inner error band around the calculation 
indicates the renormalisation scale uncertainty, the
outer band includes variations of $m_c$ and $\varepsilon$.
Within the experimental uncertainties the cross section is well
described.

\section{Diffractive photoproduction of dijets}
\begin{figure}[!thb]
\vspace*{4.3cm}
\begin{center}
\includegraphics{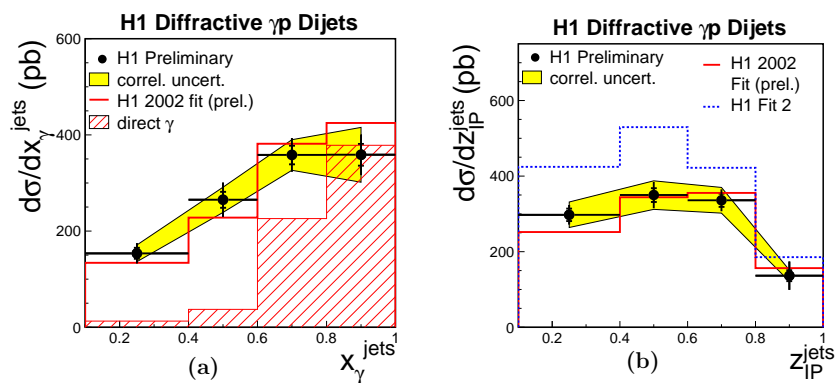}
\caption[]{Diffractive $\gamma p$ cross section as a function of
a) \xgamjets{} and b) \zpomjets{} compared with LO predictions of the RAPGAP
Monte Carlo program with parton showers to simulate higher-order
corrections. The predictions are based on LO DPDFs.}
\label{fig:gpxgam}
\end{center}
\vspace*{-1.5cm}
\end{figure}

\begin{figure}[!thb]
\vspace*{7.5cm}
\begin{center}
\includegraphics{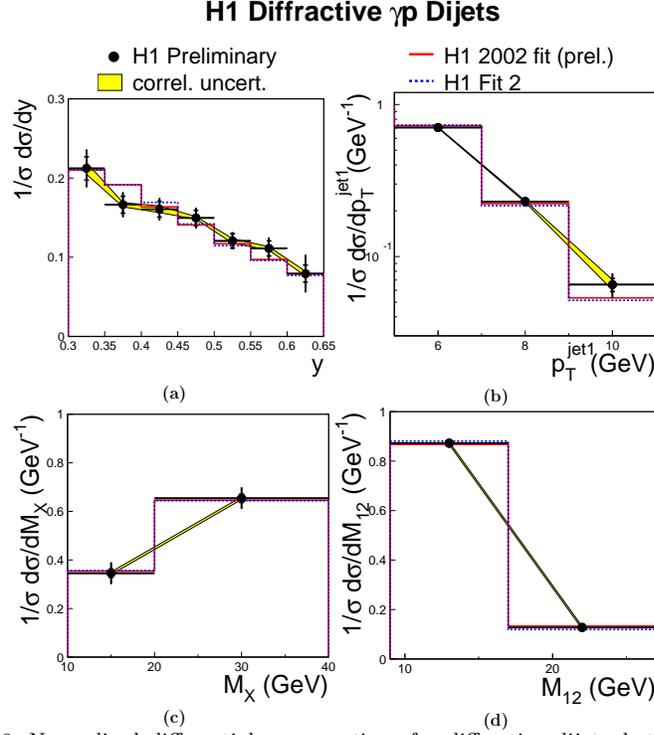}
\vspace*{2cm}
\caption[]{Normalised differential cross sections for diffractive dijet
  photoproduction compared with LO predictions of the RAPGAP
Monte Carlo program with parton showers to simulate higher-order
corrections. The predictions are based on LO DPDFs.}
\label{fig:gpother}
\end{center}
\vspace*{-0.7cm}
\end{figure}

A measurement of dijet cross sections in diffractive photoproduction
has been presented by H1 in~\cite{gpjets} for the kinematic range
$Q^2<0.01$~GeV, $\xpom<0.03$, $\etjet(1)>5$~GeV and $\etjet(2)>4$~GeV 
where jets are identified using the inclusive $k_T$ cluster 
algorithm~\cite{kt}.
In \figref{fig:gpxgam} and \ref{fig:gpother}, the measured 
distributions are compared with RAPGAP predictions with parton showers
enabled to simulate higher-order corrections.
The contribution of direct photon processes  ($\xgam=1$ at the
generator level) 
amounts to approximately half of the cross
section and is shown as the hatched
histogram in \figref{fig:gpxgam}a. The cross section is well described
by the prediction based on the recent `H1 2002 fit' PDFs throughout the 
measured $\xgam$ range which covers regions that are
dominated by either direct or resolved processes. 
The `H1 fit 2'
prediction overestimates the rate by a factor $\approx 1.4$ (\figref{fig:gpxgam}b).
The normalised cross section is shown as a function of $y$, $\ptjetone$,
$M_X$ and $\mjj$ in
\figref{fig:gpother}. The shapes of all measured distributions are well
described by both predictions.

The results for diffractive dijet production in DIS and
photoproduction can be compared to examine a possible suppression
in photoproduction relative to DIS which is expected by gap survival
probability models.
The ratio of LO prediction to data, where
the RAPGAP Monte Carlo prediction with parton showers is used
in both DIS and photoproduction, is found to be a factor 
$1.3\pm 0.3$~(exp.) larger in photoproduction.
The uncertainty is estimated from the total experimental errors of
the two measurements only. 
The 
factor is independent of the DPDFs used in the comparison.
This LO suppression is not significant at the present level of
precision
and there is no evidence that it differs between direct and resolved
photon processes.

\section{Conclusions}
NLO QCD calculations based on NLO diffractive parton densities
and assuming QCD factorisation are compatible within the experimental
and theoretical uncertainties with measurements of diffractive dijet 
and D$^*$ production in DIS. The concept of QCD factorisation holds
in diffractive DIS.
LO Monte Carlo QCD calculations based on LO diffractive parton
densities and using parton showers to simulate higher-order
corrections are in agreement with measurements of diffractive dijet 
production in DIS and photoproduction within the uncertainties.
A leading order suppression factor
for diffractive dijet photoproduction
relative to the same process in DIS of $1.3\pm 0.3$~(exp.) is found.
This factor does not deviate from unity at the present level of
precision and it does not differ between direct and resolved photon 
processes.

\end{document}